\documentclass[11pt,a4paper]{article}
\usepackage[a4paper, left=2cm, right=2cm, top=2cm, bottom=2cm]{geometry}
\usepackage{graphicx}
\usepackage[english]{babel}
\usepackage[T1]{fontenc}
\usepackage{lmodern}
\usepackage{floatflt}
\usepackage{longtable}
\usepackage{fancybox}
\usepackage[hyphens,obeyspaces,spaces]{url}
\usepackage{color}
\usepackage{amssymb}
\usepackage{blindtext}
\usepackage{parskip}
\usepackage{pdflscape}
\usepackage{amsmath}
\usepackage{natbib}
\usepackage{array}
\usepackage{setspace}
\usepackage{capt-of}
\usepackage{makeidx}
\usepackage{hyperref}
\usepackage{doi}
\usepackage[printonlyused]{acronym}
\usepackage{booktabs}
\usepackage[nottoc,notlot,notlof]{tocbibind}
\usepackage[stable]{footmisc}
\usepackage{rotating}
\usepackage{ulem}
\usepackage{lineno}
\usepackage{parskip}

\usepackage{chngcntr}
\counterwithout{figure}{section}
\counterwithout{table}{section}

\addto{\captionsenglish}{}
\begin{document}
%\maketitle

\section*{From trees to rain: Enhancement of cloud glaciation and precipitation by pollen}
Jan Kretzschmar$^{1,*}$, Mira Pöhlker$^{2,1}$, Frank Stratmann$^2$, Heike Wex$^2$, Christian Wirth$^{3,4}$, and Johannes Quaas$^{1,3}$\\
1 Leipzig University, Institute for Meteorology, Stephanstr.\,3, 04103\,Leipzig, Germany\\
2 Leibniz-Institute for Tropospheric Research, Permoser Str.\, 15, 04318\,Leipzig, Germany\\
3 German Centre for Integrative Biodiversity Research (iDiv) Halle-Jena-Leipzig, Puschstr.\,4, 04103\,Leipzig, Germany\\
4 Max-Planck-Institute for Biogeochemistry, Hans-Knöll-Str. 10, 07745 Jena, Germany\\ 
$^{*}$ Corresponding author: Jan Kretzschmar (jan.kretzschmar@uni-leipzig.de)
\vspace{0.5cm}\\
Classification: Physical Sciences, Atmospheric Sciences\\
Keywords: pollen, ice nucleating particles, cloud ice fraction, precipitation 

\vspace{0.5cm}
\paragraph*{Abstract} \mbox{} \\
The ability of pollen to enable the glaciation of supercooled liquid water has been demonstrated in laboratory studies; however, the potential large-scale effect of trees and pollen on clouds, precipitation and climate is pressing knowledge to better understand and project clouds in the current and future climate. Combining ground-based measurements of pollen concentrations and satellite observations of cloud properties within the United States, we show that enhanced pollen concentrations during springtime lead to a higher cloud ice fraction. We further establish the link from the pollen-induced increase in cloud ice to a higher precipitation frequency. In light of anthropogenic climate change, the extended and strengthened pollen season and future alterations in biodiversity can introduce a localized climate forcing and a modification of the precipitation frequency and intensity.

\paragraph*{Significance statement} \mbox{} \\
For ice crystals to form in a temperature range between 0$\,^\circ$C and –38$\,^\circ$C, the presence of so-called ice nucleating particles is needed. Laboratory studies have shown that pollen grains posses such ice nucleating properties. In this study, we show that the ice nucleating effect of pollen is observable in satellite observations and that the resulting higher cloud ice content causes an enhanced precipitation frequency.

\vspace{0.5cm}
\paragraph*{Introdution} \mbox{} \\
Clouds may consist of supercooled liquid water in a temperature range between 0$\,^\circ$C and –38$\,^\circ$C. Whether or not the cloud glaciates in this temperature regime depends on the availability of a subset of atmospheric aerosol particles, the ice nucleating particles (INP), which enable the freezing of supercooled liquid droplets or haze particles via heterogeneous freezing mechanisms. The thermodynamic state of clouds is of particular importance for their propensity to form rain. The vast majority of rain events originate from ice clouds \citep{Mulmenstadt2015,Field2015}. It has been demonstrated that pollen are among the aerosol species that may serve as INP \citep{Diehl2001,Diehl2002,VonBlohn2005,Pummer2012,Augustin2013,Hader2014}. While the contribution of pollen to the glaciation of mixed-phase clouds is thought to be rather small on a global scale compared to other INP like dust, they can nevertheless play an important role on regional and seasonal scales \citep{Pummer2012,Despres2012}. 
\par
Laboratory studies, investigating the freezing temperature of supercooled water droplets with an embedded pollen grain indicate that most pollen induce freezing in a temperature range between -15$\,^\circ$C and -25$\,^\circ$C \citep{Gute2020}. To be at this temperature regime, pollen have to be transported into higher layers of the atmosphere where such temperatures can be reached. Ground-based lidar observations have revealed that layers containing pollen can indeed reach those temperature regimes, likely being lifted to those altitudes by vertical mixing due to convection and turbulence, or due to large-scale uplift  \citep{Bohlmann2021}. 
Due to their relatively large size, whole pollen grains have a limited residence time in the atmosphere before they are removed by gravitational settling \citep{Phillips2008a,Steiner2015}. However, it is not the whole pollen grain itself but rather macromolecules on their surface that act as INP \citep[e.g.,][]{Pummer2012, Augustin2013, Steiner2018, Mikhailov2021, Burkart2021}. Under humid conditions, pollen grains rupture into so-called subpollen particles. Due to their smaller size, these particles have a longer residence time in the atmosphere and can reach higher altitudes than a whole pollen grain, increasing their ability to act as INP in the mixed-phase temperature regime \citep{Pummer2012,Steiner2015,Seifried2021}.
\par
\begin{figure}[t]
\begin{center}
\includegraphics[width=.80\textwidth]{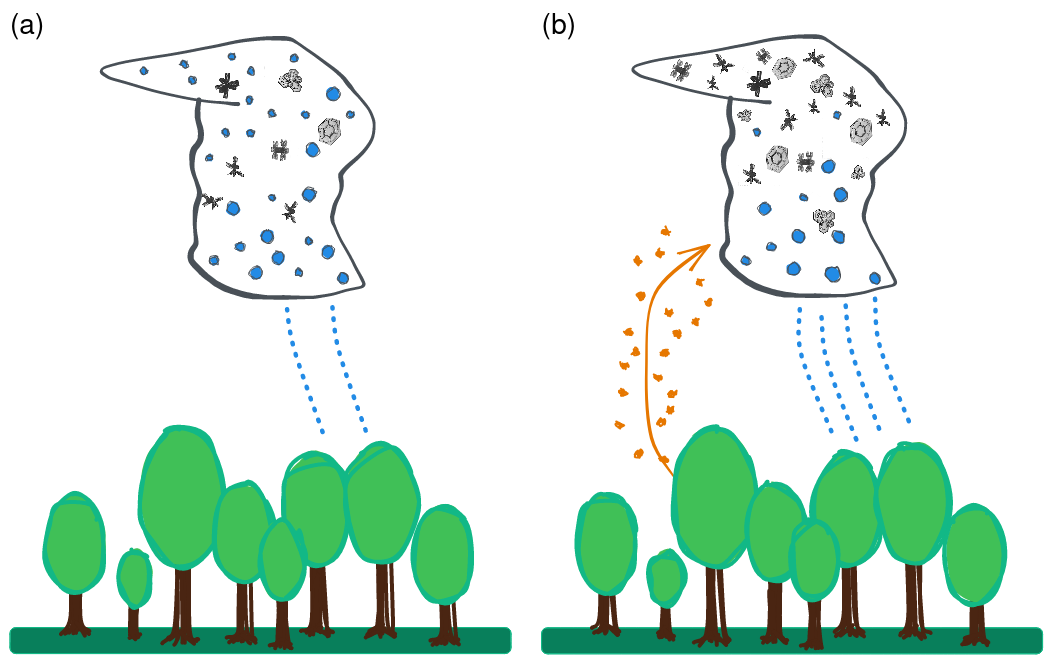}
\end{center}
\caption{Schematic depiction of the proposed glaciating effect of pollen on mixed-phase clouds. During pollen season (b), cloud ice fraction is increased compared to situations with low pollen concentration (a), consequently leading to an increase in rain frequency.}\label{pollen_cloud_diagramm}
\end{figure}

While the ice activity of pollen is widely researched in laboratory studies, so far there is no evidence on the role of pollen for cloud glaciation from observations at a large scale. In this study, we, therefore, aim at assessing whether the effect of pollen on the cloud ice fraction in the heterogeneous freezing temperature regime can be quantified using ground-based pollen concentration and satellite observations of clouds and whether this induces an increased precipitation frequency in mixed-phase clouds as illustrated in Fig. \ref{pollen_cloud_diagramm}.

\begin{figure}[t]
\begin{center}
\includegraphics[width=.90\textwidth]{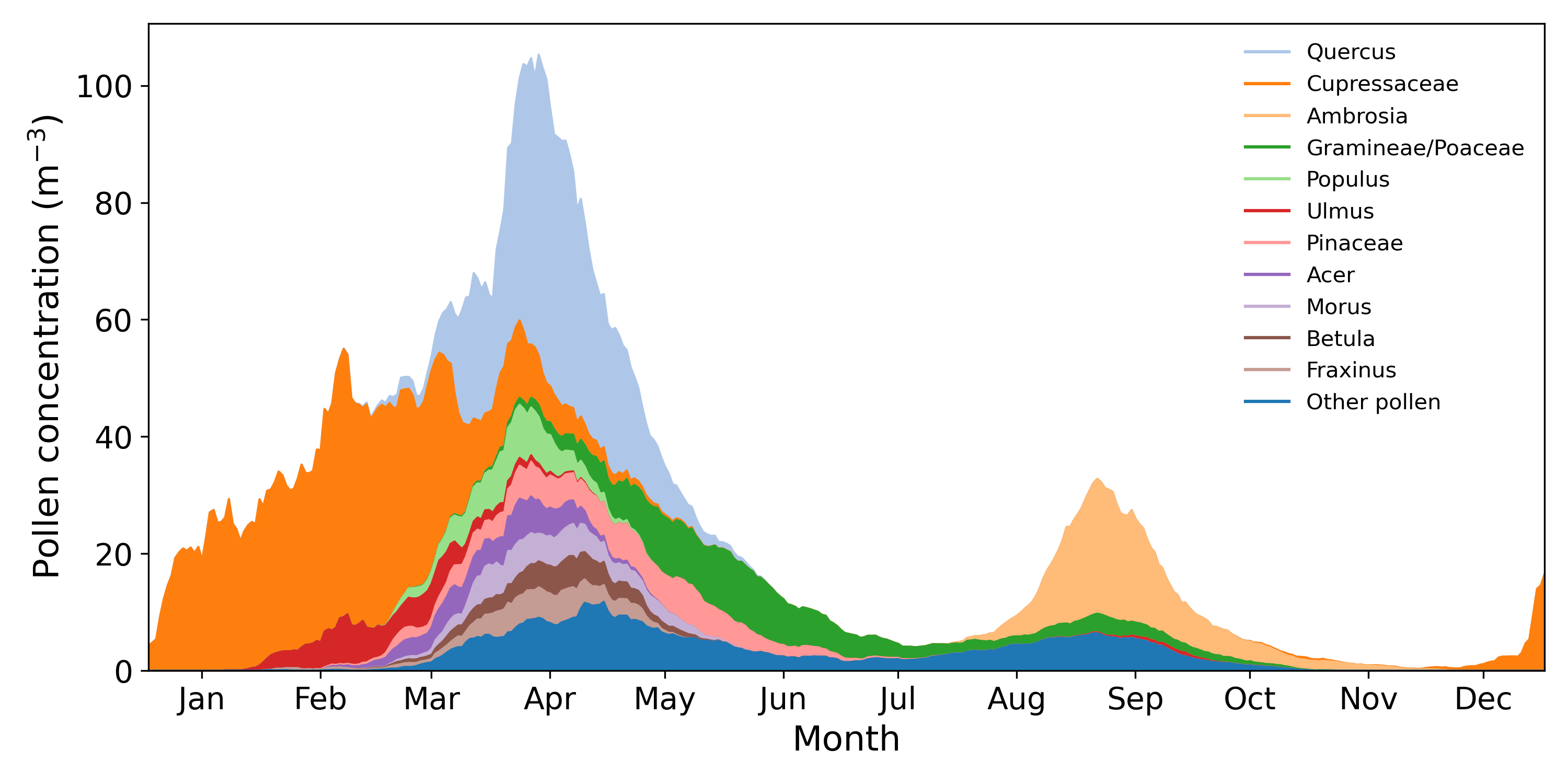}
\end{center}
\caption{Median of pollen concentration for multi-year averaged seasonal cycle at all pollen stations used in this study. A seven-day running mean was applied to smooth the time series. The eleven most frequently counted pollen taxa are shown at a genus or family resolution. All other pollen taxa are subsumed under "Other pollen".}
\label{pollen_station_count}
\end{figure}

\paragraph*{Results} \mbox{} \\
As a proxy for pollen concentration in the atmosphere, we employ ground-based pollen concentration observations, collected from stations within the United States (US; the location of pollen stations used is given in Fig. \ref{map_icefrac_dif}). Data from these pollen stations are disseminated by the National Allergy Bureau (NAB), which is part of the American Academy of Allergy, Asthma and Immunology (AAAAI). In this study, we use pollen concentrations collected by more than 50 surface stations from 2007 to 2016. These stations use volumetric air samplers from which the daily pollen concentrations are derived. \cite{Wozniak2017} showed that two distinct maxima in pollen concentration can be identified in the US, one during spring and one in fall, which can be consistently identified across all regions of the US. In the following, we will focus on the springtime (March-April-May, MAM) maximum,  to which mostly deciduous broadleaf trees contribute. As shown in Fig. \ref{pollen_station_count}, the emission of different pollen taxa in this period is strongly temporally correlated, making it difficult to disentangle the effect of a single pollen taxon. Therefore, we decided to simplify our analysis by only using the total pollen concentration observed at the respective stations as a proxy for pollen concentration in the atmosphere.
\par
In the following, we compare cloud properties between high and low pollen conditions. Low pollen conditions are defined as situations where pollen concentration, as measured at the surface, is less than $10\,\text{m}^{-3}$, whereas pollen concentration is considered to be high when pollen concentrations of larger than $60\,\text{m}^{-3}$ are observed. The upper threshold of $60\,\text{m}^{-3}$ is representative of the mean in total pollen concentration during MAM in the USA (see Fig. \ref{pollen_station_count}).
\par
To explore the effects of pollen on cloud glaciation, we first employ data from passive satellite remote sensing, and in a second step, also active remote sensing is used. The advantage of the former is the much larger statistical sample due to the wide swath. The advantage of the latter is a less uncertain determination of the cloud thermodynamic phase. As passive remote sensing dataset, we use ice fraction as a function of cloud-top temperature derived from Moderate Resolution Imaging Spectroradiometer (MODIS) Level-2, Collection 6.1 at a horizontal resolution of 1\,km \citep{Platnick2017}. An evaluation of the employed MODIS retrieval algorithm for cloud phase shows good agreement with cloud phase derived from the spaceborne CALIPSO lidar \citep{Marchant2016}, providing confidence in the retrieved ice fraction. Around each pollen station, we consider the MODIS retrievals within a circle with a radius of 100\,km, inside which we assume the pollen concentration to be represented by the one observed at the surface pollen station. Sensitivity studies demonstrated that the results are independent of the exact choice of radius (not shown). 
\par
\begin{figure}[t]
\begin{center}
\includegraphics[width=.90\textwidth]{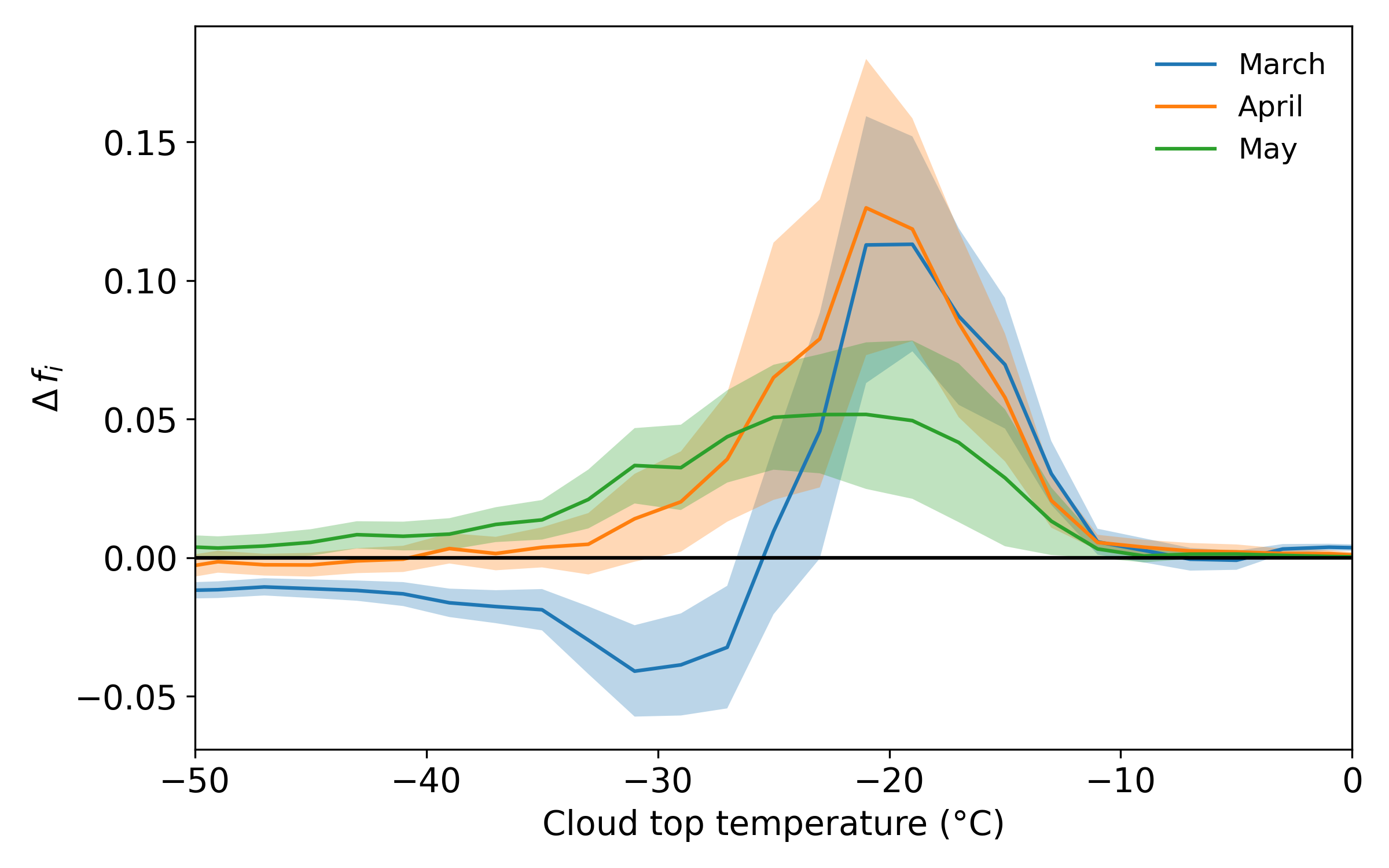}
\end{center}
\caption{MODIS-derived mean difference in ice fraction $\Delta f_\text{i}$ between high and low pollen concentrations as a function of cloud-top temperature. $\Delta f_\text{i}$ is binned every 2 K and data are averaged for ten years from 2007 to 2016. MODIS data are sampled within a radius of 100 km around each pollen station. Low pollen cases are defined as having a total pollen concentration at the station of less than 10 m-3 and high pollen cases are defined as having a total pollen concentration at the station of greater than 60 m-3. Shaded areas indicate the 95\,\% confidence interval of the mean as estimated from bootstrapping using a sample size of 10000.}\label{cloud_phase_dif}
\end{figure}
Fig. \ref{cloud_phase_dif} shows the mean difference in ice fraction as diagnosed from the MODIS dataset for cases with high and low pollen concentrations. We find a maximum in the difference between high and low pollen conditions at around -20$\,^\circ$C, which is most strongly expressed in March and April and is slightly reduced in May. For all months, the positive difference in ice fraction is statistically significant for a cloud top temperature range between -17$\,^\circ$C and -23$\,^\circ$C. This increased ice fraction is in good agreement with the freezing temperature of a water-embedded pollen grain between -15$\,^\circ$C and -25$\,^\circ$C, reported in laboratory studies \citep[][]{Gute2020}. We would like to remark that for temperatures greater than -13 °C, the cloud phase retrieval of MODIS gives large weight towards the liquid phase (supplemental material in \cite{Platnick2017}), so the difference between high and low pollen concentrations is close to zero in that temperature regime. 
\par
To further constrain the influence of pollen on cloud ice fraction, we subdivided the continental US into three sub-regions (western, southeastern, and northeastern US; Fig. \ref{map_icefrac_dif}). When focusing on the two regions in the eastern US, we find a temporal variation in the response of cloud ice fraction to pollen concentration. The largest difference in ice fraction for the southeastern US is observed already in March, whereas the suspected effect of pollen on ice fraction only can be seen for April and May in the northeastern US. This delayed response in ice fraction coincides with the later start of the pollen season in higher latitudes \citep{Lo2019}, strengthening the causal relationship between the presence of pollen and the increased cloud ice fraction.
\par
As meteorological conditions strongly change during springtime, we evaluated whether the above-reported signal in ice fraction might stem from a temporal covariability of pollen and meteorology. We first evaluated whether there is a general temporal trend in cloud ice fraction during springtime by comparing the beginning to the end of all springtime months. We find that cloud ice fraction decreases during each month in the mixed-phase cloud temperature regime (see Fig. \ref{month_start_end_dif}). From Fig. \ref{pollen_station_count} we see that pollen concentration on US average is increasing in March, rather constant in April and decreasing in May. The fact that there is a temporal trend of pollen concentration in some months implies that filtering for low or high pollen concentration will lead to an implicit temporal sampling. For example, low pollen concentration occur in the beginning and high pollen concentration occur during the end of March. In combination with the observed decreases in ice fraction during March, this implicit temporal sampling contributes to the negative cloud ice fraction differences in March for temperatures less than \mbox{-25$\,^\circ$C}. For this reason, the signal of pollen would be even more strongly expressed without the alteration in ice fraction due to changing meteorological conditions in this month. The effect of implicit temporal sampling flips sign in May, as there is now a decreasing trend in pollen emission towards the end of the month, so part of the increased cloud ice fraction in Fig. \ref{cloud_phase_dif} might be related to this implicit temporal sampling.
\par
\begin{table}[t]
\centering \begin{tabular}{l|c|c|c}
\toprule
$\Delta_{\text{rel}}$ (high - low pollen) &  March (\%) &  April (\%) &  May (\%) \\
\midrule
Black carbon (hydrophilic)   &         8.0 &         0.8 &      -1.7 \\
Organic matter (hydrophilic) &        15.0 &         3.1 &      -6.9 \\
Black carbon (hydrophobic)   &        -0.8 &        -2.0 &       5.7 \\
Organic matter (hydrophobic) &         4.3 &        -1.8 &       1.2 \\
Sulphate                     &        -7.3 &        -2.6 &       3.0 \\
Dust                         &       -26.6 &       -13.3 &      12.2 \\
Sea salt                     &        -6.8 &         3.5 &      14.2 \\
\bottomrule
\end{tabular}

\caption{Relative difference between high and low pollen concentrations in column load for aerosol types in the CAMS aerosol reanalysis.}\label{tab_aerosol_pollen}
\end{table}
Besides pollen, other aerosol species like dust may act as INP in the temperature range where pollen are ice-active. For that reason, we further investigated whether the presence of pollen is spatiotemporally correlated with other aerosols. As there is only a limited amount of long-term observations of aerosol available in the vicinity of pollen stations, we use information on the presence of aerosol from the Copernicus Atmospheric Monitoring Service \citep[CAMS;][]{Inness2019} aerosol reanalysis. CAMS is based on the European Centre for Medium-Range Weather Forecast (ECMWF) Integrated Forecast System model, which has been extended to model the emission, transport and removal of aerosols and trace gases. The modelled aerosol optical depth (AOD) in CAMS is further constrained by assimilating spaceborne observation of AOD from Envisat's Advanced Along-Track Scanning Radiometer \citep[AATSR, 2003-2012][]{Popp2016} and MODIS \citep[2003-present;][]{Levy2013}. CAMS has proven to perform well in comparison with ground-based observation of AOD and also in comparison with the MERRA-2 aerosol reanalysis \citep{Gueymard2020}. In Tab. \ref{tab_aerosol_pollen}, we show the relative change in aerosol column load between high and low pollen concentration for different aerosol species modelled in CAMS. We find that dust load is reduced by up to 20\,\% in March and April, which are the month where we found the strongest susceptibility of cloud ice fraction to pollen. Dust load in the atmosphere over continental US has a seasonal cycle with a maximum during summer. During springtime, dust load is already increasing, but as indicated in Fig. \ref{month_start_end_dif}, cloud ice fraction is decreasing during springtime, so any effect of the increased dust load is more than compensated by the seasonal modification of ice fraction. As stated above, the effect of pollen on ice fraction is more than able to compensate the seasonal alteration in ice fraction, further highlighting their strong ice nucleating properties. Other aerosol species that might act as INP like black carbon and organic matter only show a rather weak correlation with pollen concentrations, further strengthening the relationship between ice fraction and pollen concentration.
\par
\begin{figure}[t]
\begin{center}
\includegraphics[width=.70\textwidth]{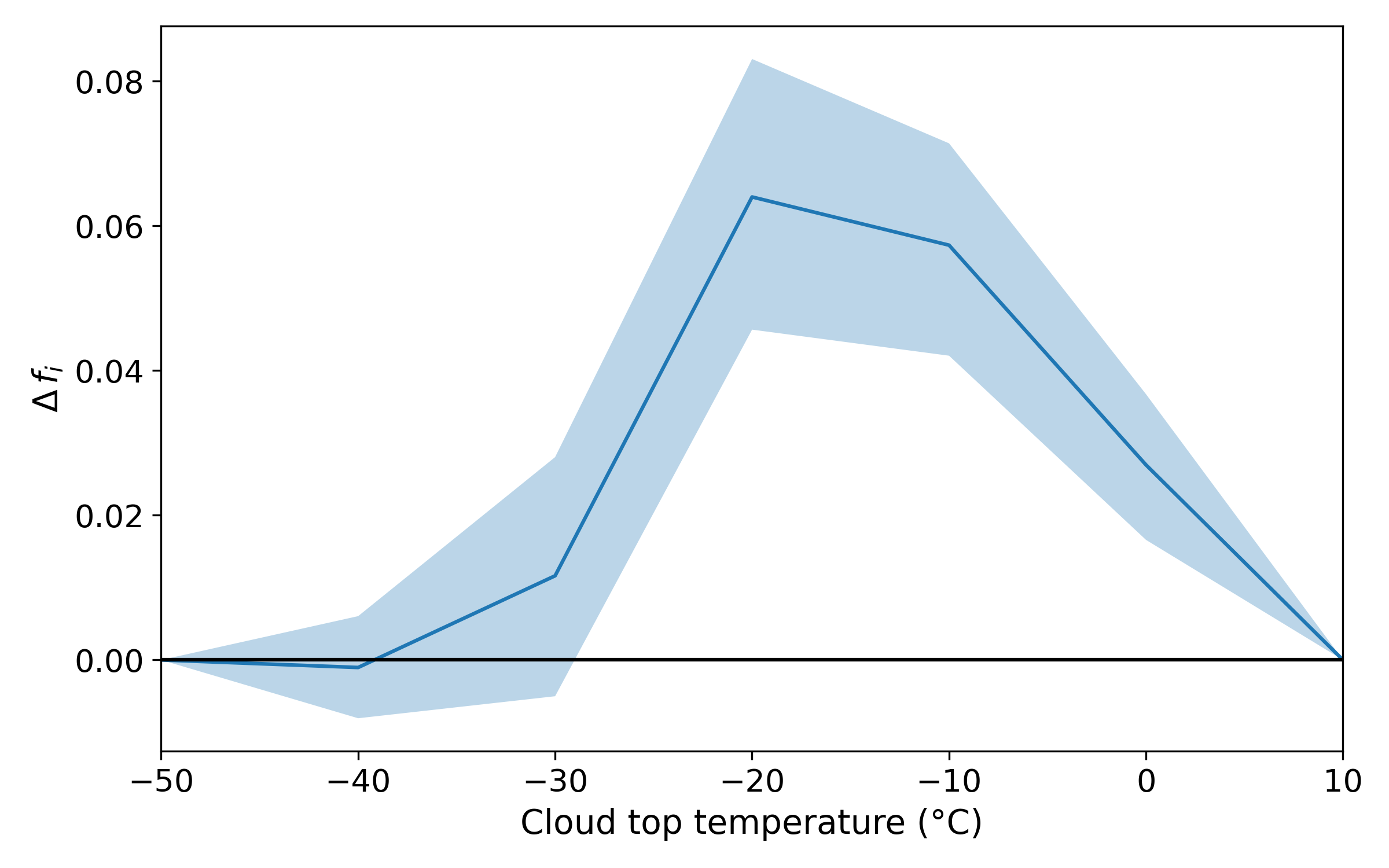}
\end{center}
\caption{As Fig.\,\ref{cloud_phase_dif}, but from DARDAR active satellite remote sensing retrievals, sampled within a radius of 200\,km around each pollen station, using 10\,K bins in cloud temperature, and considering the March-April-May period.}\label{cloud_phase_dif_dardar}
\end{figure} 
To verify the results from MODIS, we additionally employed data from active satellite remote sensing. We use DARDAR \citep[raDAR/liDAR,][]{Delanoe2010}, which combines information from a spaceborne cloud radar \citep[CloudSat; ][]{Stephens2002} and cloud lidar \citep[CALIPSO; ][]{Winker2003}. In particular, we utilize the so-called DARDAR\_MASK product, which contains information on atmospheric features like the phase state of hydrometeors as well as the presence of atmospheric aerosols. The difference in cloud ice fraction as a function of cloud top temperature for DARDAR is shown in Fig. \ref{cloud_phase_dif_dardar}. We again find an increased ice fraction which peaks at -20$\,^\circ$C. While the maximum is in accordance with the MODIS-derived difference in ice fraction, the DARDAR-derived positive difference in ice fraction extends towards warmer temperatures up to the freezing point. The difference to MODIS can be related to the more sensitive cloud phase retrieval of DARDAR with respect to ice clouds especially at temperatures greater than -10$\,^\circ$C. According to laboratory studies, pollen are not considered to be strongly ice active at such elevated temperatures, so this signal can potentially be related to aerosol species like bacteria and/or fungal spores, which are ice active in this temperature regime \citep[e.g.,][]{Murray2012,Kanji2017} or to aforementioned cross-correlation with meteorological conditions.
\par
Using the ability of DARDAR to penetrate through optically thick clouds and to retrieve information on cloud and hydrometeor properties almost down to the surface, we can assess whether an effect of the increased ice fraction due to the presence of pollen on precipitation can be identified. As most precipitation over the continents stems from the ice phase \citep{Mulmenstadt2015}, it is expected that the enhanced cloud ice fraction in response to a higher atmospheric pollen concentration leads to an increase in the fraction of clouds that precipitate. While for low pollen cases, the fraction of precipitating clouds is $9.19\,\% \pm 0.37\,\%$, it increases to $11.82\,\% \pm\,0.19\,\%$ for high pollen cases, where the given uncertainty is the 95\,\% confidence interval of the mean derived from bootstrapping with a sample size of 10000. We remark that not all of this increase is causally linked to the increase in ice-containing clouds. Further causes are a temporal correlation between the seasonal shift in precipitation frequency and pollen concentration as discussed for cloud ice fraction. Also, the opposite causality exists: precipitation affects pollen concentration in the atmosphere. While precipitation reduces the amount of aerosol in the atmosphere through wet deposition, it has also been demonstrated that pollen concentrations can even increase before and during rainfall events \citep{Kluska2020}. A reason for this is that pollen can get lifted by higher wind speeds before and during rainfall events. 
\par
To be able to directly relate alteration in cloud ice fraction between high and low pollen concentrations to alteration in precipitation frequency, we calculated how this change translates into changes in precipitation frequency, given the cloud distribution by temperature and the change in precipitation probability (see Methods). We find, as expected, a smaller effect of pollen on the fraction of precipitating clouds from $9.53\,\% \pm 0.02\,\%$ for low pollen cases to $10.13\,\% \pm\,0.01\,\%$ for high pollen cases. This still is a substantial absolute increase of 0.6\,\% more frequent rain due to the alteration in cloud glaciation over the continental USA during springtime. 

\section*{Conclusion}
We have shown that on a regional scale, pollen can have a significant ice nucleating effect, in particular during springtime when large amounts of pollen are emitted. This is in contrast to the assumed global effect of pollen, which is thought to be low as reported by \cite{Hoose2010a} from a study with a global climate model. This study did not consider the effect sub-pollen particles on cloud microphysics, which consequently will lead to an underestimated contribution of pollen to the amount of ice nucleating particles. The importance of sub-pollen particles has been highlighted in modelling studies for their effect as CCN \citep{Steiner2018} and INP \citep{Werchner2023} and underlines the importance to include effects in sub-pollen particles into climate models to correctly estimates their effect on cloud microphysics.
\par
Anthropogenic climate change has already been shown to shift the start of springtime pollen emissions, lengthen the pollen season and increase the concentration of airborne pollen \citep{Ziska2019,Anderegg2021}. These trends will continue to manifest themselves towards the end of the century due to increased temperatures, changes in precipitation amount and frequency and the fertilizing effect of a higher CO$_2$ concentration \citep{Zhang2022}. Our results show that those changes can, at least regionally and during springtime, have a significant effect on cloud glaciation leading to an increase in precipitation frequency. The circumstance that several taxonomic groups of trees jointly produce the distinct peak in pollen production during MAM points to a potential role of biodiversity in controlling cloud glaciation and precipitation which demands further research.

\section*{Data and methods}
We employ ground-based pollen concentration data collected from stations within the United States (US) (the location of pollen stations used is given in Fig. \ref{map_icefrac_dif}). Pollen station data are disseminated by the National Allergy Bureau (NAB), which is part of the American Academy of Allergy, Asthma and Immunology (AAAAI). In this study, we use pollen concentrations collected by more than 50 surface stations in a period from 2007 to 2017. Each timeseries at a station contains pollen concentrations of up to 40 different pollen taxa, but due to a strong temporal correlation in pollen emission, we simplify the analysis by only using total pollen concentration in our analysis. We need to remark that not all stations cover the full time period of interest, but cloud properties in the vicinity of a pollen station are only sampled if information on pollen concentration is available. We chose a radius of 100\,km around a pollen station where we sample cloud properties from satellite products as we consider pollen emission in a region to be spatiotemporally homogeneous. Such conditions can to some extent be assumed for regional scales of a few hundred kilometers with similar climatic conditions and similar composition of pollen emitting plants \citep{Nowosad2015}. We nevertheless do not want to overextend this assumption, as long-range transport clearly has been shown to decorrelate observed pollen concentrations from local emissions \citep[e.g.][]{Sofiev2017}. For that reason, we evaluated the effect of using different sampling radii around the pollen stations for cloud properties from the satellites, but the ain findings in this study were independent of the employed sampling radius.
\par
We use daily data from the Moderate Resolution Imaging Spectroradiometer (MODIS) Level-2, Collection 6.1 dataset at a horizontal resolution of 1\,km, from both, the Aqua and Terra satellite. Due to the wide swath of MODIS, a large number of satellite pixels are available in the vicinity of a pollen station, which enables us to calculate cloud ice fraction f$_\text{i}$ as a function of temperature at each station and timestep. At each cloudy satellite pixel, the MODIS cloud phase retrieval indicates if a cloud is liquid or ice or if the cloud phase retrieval is uncertain. Due to the fact that the cloud phase retrieval in the MODIS dataset is dependent on information from shortwave spectral bands, we only use daytime overpasses in our analysis. Furthermore, we only consider pixels that are flagged as single-layer clouds by the MODIS retrieval to avoid uncertain retrievals in cloud top temperature and phase. From the ratio of pixels in the ice phase to the total number of cloudy pixels within the 100\,km sampling radius, we calculate f$_\text{i}$, binned by cloud top temperature at each station ($s$) and each satellite overpass/timesteps ($t$), defined as:
\begin{equation}
    \text{f}_{\text{i}}(T, s, t) = \dfrac{n_{\text{ice}}(T, s, t)} {\sum_{j} n_{j}(T, s, t)} \quad j \in  \{ \text{ice}, \text{liquid}, \text{undetermined} \} \, .
\end{equation}
Here we use a bin width of 2\,K between -50$\,^\circ$C and 10$\,^\circ$C. To compare high and low pollen cases, we calculate the mean of f$_\text{i}$ at each temperature bin among all stations and timesteps, respectively. Here, a weighted mean is employed to give more weight to situations having more cloudy pixels within the sampling radius, such that:
\begin{equation}
    \overline{\text{f}_{\text{i}}}(T) = \sum_{s, t} w(T, s, t) \,\, \text{f}_{\text{i}}(T, s, t) \quad \text{with} \quad w(T, s, t) = \dfrac{\sum_{j} n_{j}(T, s, t)}{\sum_{j,s,t} n_{j}(T, s, t)}
\end{equation}
To quantify statistical uncertainty in the difference of $\overline{\text{f}_{\text{i}}}(T)$ between high and low pollen cases, we bootstrapped the mean of high and low pollen cases using a sample size of 10000 to calculate the  95\,\% confidence interval for this difference.
\par
We additionally use information on cloud properties from DARDAR \citep[raDAR/liDAR,][]{Delanoe2010}, combining data from a spaceborne cloud radar \citep[CloudSat; ][]{Stephens2002} and cloud lidar \citep[CALIPSO; ][]{Winker2003}. In particular, we use information from the so-called DARDAR\_MASK, which contains information on atmospheric features like the phase state of clouds, precipitation and atmospheric aerosols. The fact that information on atmospheric features are sampled along the ground track of the two satellites drastically limits the number of available data points in the vicinity of a pollen station. For that reason, we only look at the entire springtime period (MAM), increased the sampling radius to 200\,km, and additionally increased the width of the temperature bins to 10\,K. As DARDAR employs information from active sensors that are independent of insolation, we additionally use nighttime overpasses to increase the number of available overpasses over pollen stations. 
\par
To be comparable to MODIS, we first have to detect cloud top in the DARDAR dataset. To avoid spurious detection of a cloud layer in DARDAR, at least four consecutive cloudy points within each vertical profile (which is equivalent to a geometrical cloud depth of at least 240\,m) have to be present. The cloud phase at cloud top is then considered in the ice fraction calculation. We only use situations where one cloud top is detected in DARDAR to be comparable to MODIS data, where we also consider only single-layer clouds. While MODIS actually infers cloud top temperature from observed radiances, cloud top temperature in DARDAR is derived from ECMWF-AUX, the spatially interpolated meteorological analysis of the European Centre for Medium-Range Weather Prediction (ECMWF). MODIS only distinguishes between liquid and ice clouds, whereas in DARDAR also mixed-phase clouds (ice+supercooled) can be detected, which we consider to be in the liquid phase. In contrast to the MODIS analysis, we do not calculate ice fraction for a single pollen station which was subsequently averaged across all pollen stations, due to the limited amount of data points along the satellite ground track in the vicinity of pollen stations. For that reason, we combine information on cloud phase from all overpasses over pollen stations and calculate $\overline{\text{f}_{\text{i}}}$ as follows:
\begin{equation}
     \overline{\text{f}_{\text{i}}}(T) =  \dfrac{n_{\text{ice}}(T)} {\sum_{j} n_{j}(T)} \quad j \in  \{ \text{ice}, \text{mixed phase}, \text{supercooled}, \text{liquid} \} \, ,
\end{equation}
where $n_j$ is the number of pixels of the respective cloud phase category.
\par
Using the ability of DARDAR to penetrate through optically thick clouds and to retrieve information on cloud and hydrometeor properties almost down to the surface, we are furthermore able to assess whether an effect of pollen on precipitation can be identified. We quantify this by comparing the fraction of precipitation clouds for high and low pollen situations. As information on whether a cloud is precipitating is derived from CloudSat, which suffers from ground clutter, we assume a cloud to be precipitating when the DARDAR\_MASK profile indicates precipitation at 500\,m above ground level. Using this information, we calculate the ratio of precipitating clouds $\overline{f_p}$ for temperatures between -50$\,^\circ$C to 10$\,^\circ$C. Besides the directly derived fraction of precipitating clouds, we recalculate this value from changes in ice fraction, enabling us to quantify the effect of modified ice fraction on precipitation when pollen are present. Any deviation for the directly calculated precipitation fraction is indicative of other processes that influence rain fraction besides the glaciating effect of pollen. The total fraction of precipitating clouds can then be calculated as follows:
\begin{equation}
    \text{f}_p(T) = \overline{\text{f}_{\text{i}}}(T) \, p_{\text{ice}}(T) + [1- \overline{\text{f}_{\text{i}}}(T)] \, p_{\text{liq}}(T) \, ,
\end{equation}
where $p_{\text{ice}}(T)/p_{\text{liq}}(T)$ is the fraction of ice/liquid clouds that precipitate at a temperature bin, which we calculated from all available DARDAR profiles in the USA from 2007 to 2016. We again employ a weighted mean to calculate the mean fraction of precipitating clouds for all temperature bins from -50$\,^\circ$C to 10$\,^\circ$C:
\begin{equation}
    \overline{\text{f}_{p,\text{calc}}} = \sum_T w(T)\,\text{f}_p(T)
\end{equation}
where $w(T)$ is the ratio of cloudy profiles to the number of cloudy profiles in that temperature bin.

\section*{Acknowledgement}
We thank all the AAAAI National Allergy Bureau pollen station data providers. The authors thank Tom Goren for valuable discussion regarding the MODIS data.

\newpage
\bibliographystyle{apadoi2}
\bibliography{library.bib}

\setcounter{table}{0}
\setcounter{figure}{0}
\renewcommand{\thefigure}{S\arabic{figure}}

\clearpage
\section*{Supplemental material}

\begin{figure*}[h!]
\begin{center}
\includegraphics[page=1, width=1.\textwidth]{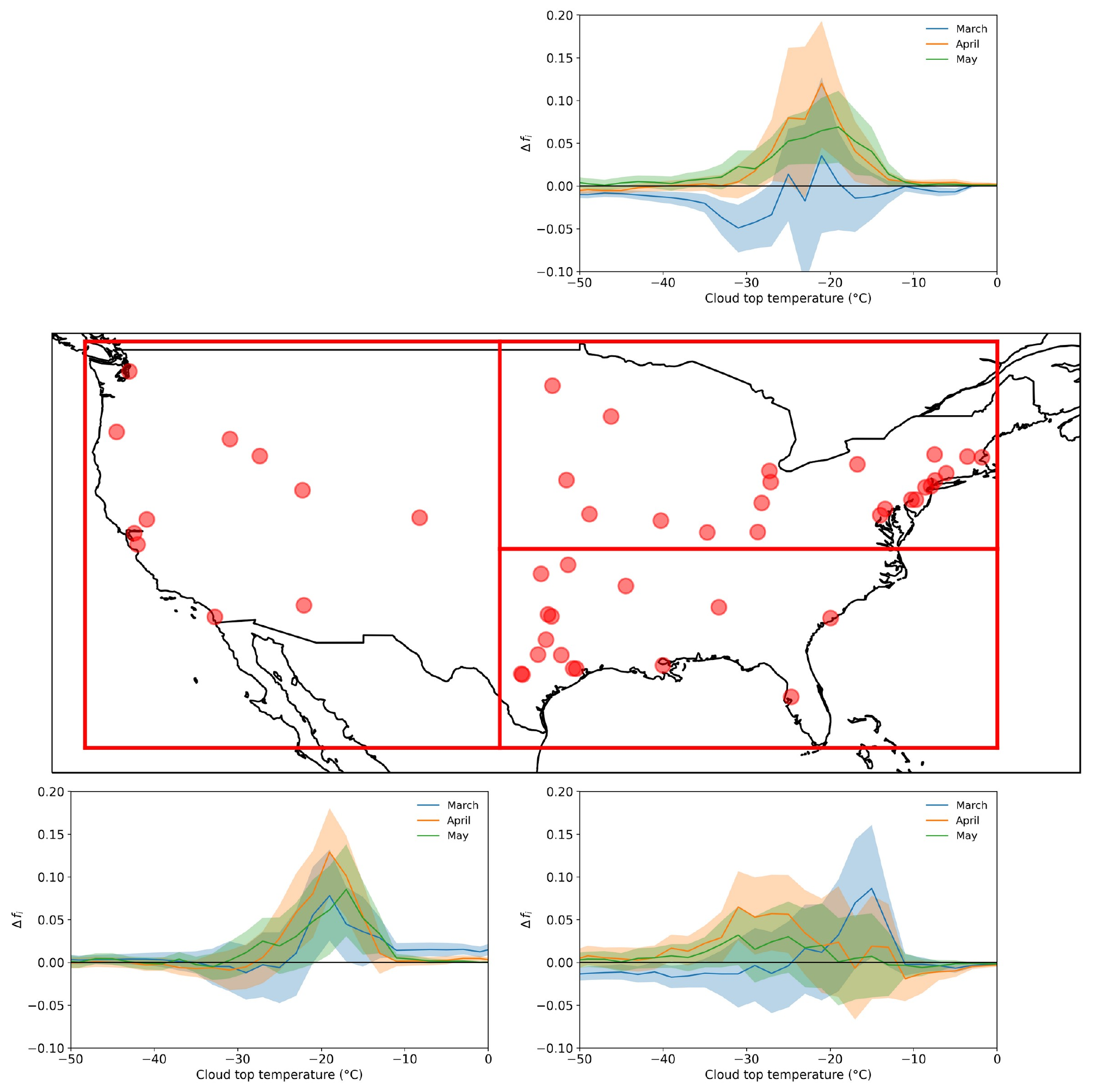}
\caption{Location of ground-based pollen stations used in the analysis and subdivision into 3 geographical regions. For each region the change in ice fraction has been calculated as in As Fig. \ref{cloud_phase_dif}.} \label{map_icefrac_dif}
\end{center}
\end{figure*}
\clearpage

\begin{figure*}[h!]
\begin{center}
\includegraphics[page=1, width=1.\textwidth]{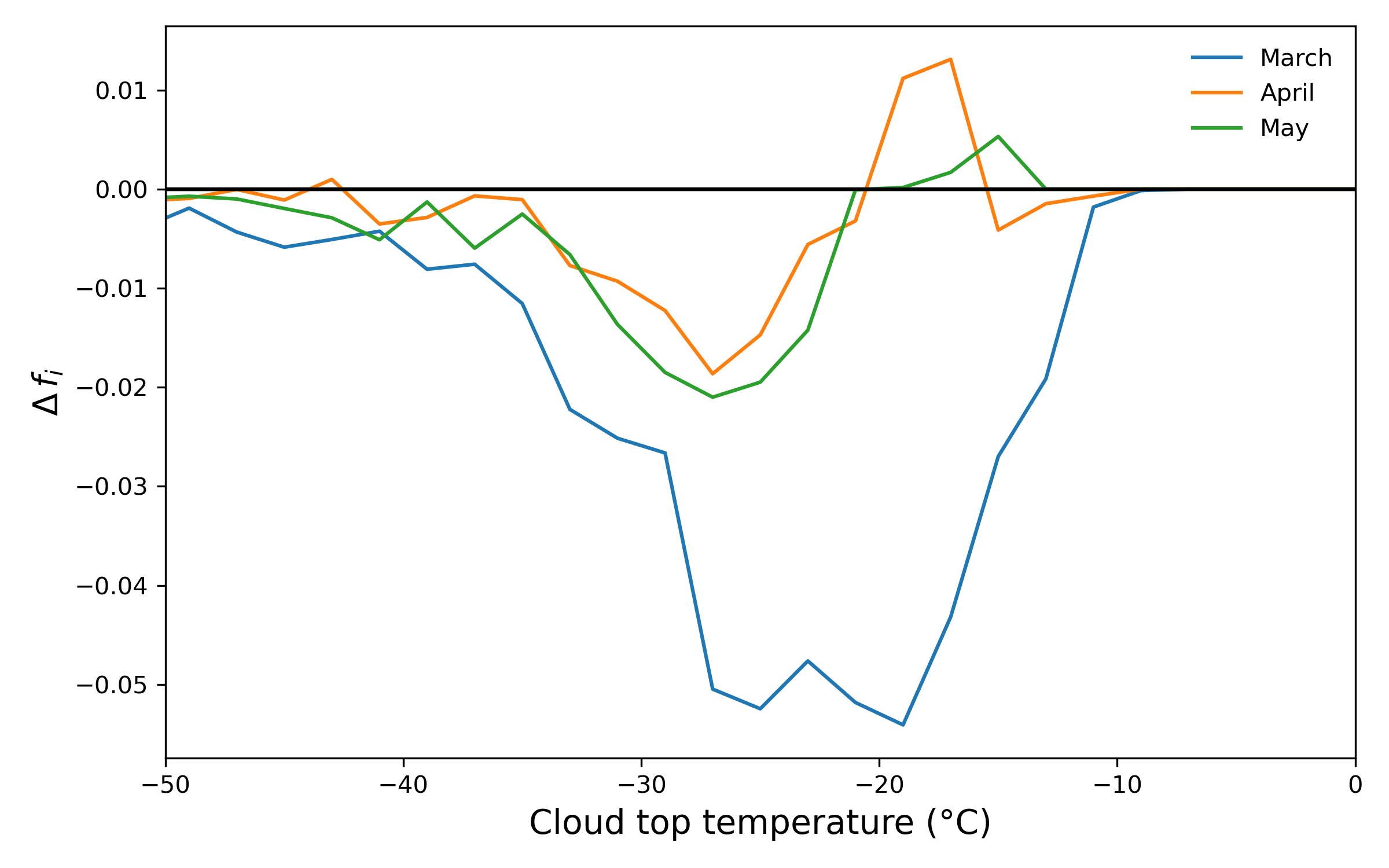}
\caption{Multi-year monthly averaged MODIS ice fraction difference $\Delta f_\text{i}$ between the last and the first ten days of the respective month, averaged over all pollen stations.}\label{month_start_end_dif}
\end{center}
\end{figure*}
\clearpage

\end{document}